\documentclass[a4paper,11pt]{article}

\usepackage{jheppub}
\usepackage[T1]{fontenc}
\usepackage{hyperref}
\usepackage{graphicx}
\usepackage{amsmath}
\usepackage{epsfig}
\usepackage{subfig}
\usepackage{xcolor}
\usepackage{natbib}
\usepackage{multirow}

\title{Inclusive productions of $J/\psi+\eta_c$ in $e^{+}e^{-}$ annihilation at Belle}
\author{Zhan Sun}
\author{,}
\author{Shu-Jian Qi}
\author{and}
\author{Ying-Zhao Jiang}
\affiliation{School of Physics and Mechatronics Engineering, Guizhou Minzu University,\\
Guiyang 550025, People’s Republic of China.}
\emailAdd{sunzhan{\_}hep@163.com}
\emailAdd{20240702001017@stu.gzmu.edu.cn}
\emailAdd{yzjiang@163.com}

\abstract{We study $e^{+}e^{-} \to J/\psi+\eta_c+X$ at next-to-leading order (NLO) in $\alpha_s$ within the nonrelativistic QCD (NRQCD) framework to assess the color-octet (CO) mechanism at low energies. QCD corrections significantly enhance both color-singlet (CS) and CO leading-order results, with non-negligible QED-diagram contributions to the CS channel. At $\Upsilon(4S)$, CO terms increase the inclusive cross section by $20\%$--$30\%$, and the enhancement grows rapidly with $\sqrt{s}$, yielding an energy dependence distinct from CS predictions and providing a sensitive test of NRQCD. For the central LDME choice, including $\psi(2S)$ feed-down further amplifies the NRQCD prediction by about $40\%$ at $\Upsilon(4S)$. With Belle's reconstruction of $J/\psi \to \mu^+\mu^-$ and six $\eta_c$ decay channels, the process shows promising potential for observation.}


\begin{document}

\maketitle

\bibliographystyle{JHEP}

\section{Introduction}\label{intro}

The study of heavy quarkonium production mechanisms has long been a central topic in high-energy physics. Within the nonrelativistic QCD (NRQCD) factorization framework~\cite{Bodwin:1994jh,Petrelli:1997ge}, significant success has been achieved in describing the transverse-momentum ($p_T$) differential cross sections of charmonia and bottomonia—such as $J/\psi$, $\psi(2S)$, $\eta_c$, $\chi_c$, and $\Upsilon$—in hadroproduction~\cite{Braaten:1994vv,Cho:1995vh,Cho:1995ce,Ma:2010vd,Zhang:2014coi,Butenschoen:2014dra,Han:2014jya,Zhang:2014ybe,Gong:2013qka,Gong:2012ug}. In particular, in the moderate-to-high $p_T$ region, the inclusion of color-octet (CO) contributions significantly improves the agreement between theoretical predictions and experimental data over the color-singlet (CS) model alone. Moreover, recent investigations of $J/\psi$ electroproduction in deeply inelastic $ep$ scattering at HERA further indicate that the NRQCD predictions provide a much better description of the data than those based solely on CS contributions~\cite{Sun:2017nly,Sun:2017wxk,Lansberg:2019adr,Boer:2024ylx}. Nevertheless, recent studies have pointed out that at colliders such as RHIC and LHC, NRQCD predictions for integrated $p_T$ cross sections exhibit clear discrepancies with experimental data~\cite{Feng:2015cba,Andronic:2015wma}. Specifically, the observed energy dependence is reasonably described by the leading-order (LO) CS prediction, whereas the inclusion of QCD corrections may lead to unphysical behavior. Moreover, in this kinematic region, the CO contributions are not dominant and are largely overshadowed by the CS component. This observation contradicts the conventional expectation that NRQCD remains valid even at low energies~\cite{Cooper:2004qe}. Indeed, the review~\cite{Andronic:2015wma} interprets this as a potential indication of NRQCD breakdown in the low-energy domain. Consequently, further investigation into the validity of NRQCD factorization at low energies is of considerable theoretical importance.

Compared with hadroproduction, electron-positron annihilation offers a cleaner environment with significantly reduced backgrounds. The center-of-mass energy of B-factories is concentrated around the $\Upsilon(4S)$ resonance, which lies in the low-energy region, thus providing an ideal platform for testing the NRQCD validity in this regime. Heavy quarkonium production in $e^+e^-$ annihilation can be broadly classified into inclusive and exclusive processes. The inclusive process $e^+e^- \to J/\psi + X$ \cite{Yuan:1996ep,Cho:1996cg,Baek:1998yf,Schuler:1998az,Kiselev:1994pu,Liu:2003zr,Zhang:2006ay,Ma:2008gq,Gong:2009ng,Gong:2009kp,He:2007te,Leibovich:2007vr,Jia:2009np,Brambilla:2010cs,Shao:2014rwa,Gong:2019rpd,Chen:2022qli,Huang:2024zni,Li:2026zsu,Jiang:2026ieb,Zhang:2009ym,He:2013ka,He:2009uf,Li:2014fya,Lee:2020dza,Fleming:2003gt,Hagiwara:2004pf,Kang:2004zj} can be further divided into the $J/\psi + c\bar{c}$ channel and the $J/\psi + X_{\text{non-}c\bar{c}}$ channel. For the former, a complete QCD next-to-leading order (NLO) study within the NRQCD framework is still lacking, as existing calculations have been restricted to the CS contributions. For the latter, the contributions from CO states can be substantial \cite{Zhang:2009ym}, even saturating or exceeding the current experimental upper limits, which strongly suggests that the corresponding long distance matrix elements (LDMEs) are severely constrained by data. Recent progress via small-$x$ resummation has partially alleviated this tension, but the issue persists \cite{Chen:2022qli}. At present, it therefore remains premature to draw a definitive conclusion on the role of the CO mechanism in inclusive $J/\psi$ production in $e^+e^-$ annihilation. For exclusive processes \cite{Liu:2002wq,Bodwin:2002fk,Braaten:2002fi,Bodwin:2002kk,Luchinsky:2003yh,Hagiwara:2003cw,Ma:2004qf,Liu:2004ga,Zhang:2005cha,Braguta:2005kr,Bodwin:2006yd,Braguta:2007ge,Gong:2008ce,Braguta:2008tg,Elekina:2009wt,Fan:2012dy,Li:2013otv,Sang:2023liy,Huang:2023pmn,Bhatnagar:2024ykb,Bondar:2004sv,Bodwin:2006dm,Choi:2007ze,Bodwin:2007ga,Bodwin:2007fz,Gong:2007db,Ebert:2008kj,Jia:2008ep,Bodwin:2010fi,Mengesha:2011pu,Dong:2012xx,Bodwin:2014dqa,Sun:2018rgx,Feng:2019zmt,Negash:2018cuk,Zeng:2021hwt,Sun:2021tma,Huang:2022dfw,Ebert:2006xq,Zhang:2008ab,Wang:2025sbx,Li:2025mng,Chen:2025qgy}, $e^+e^- \to J/\psi+J/\psi$, although easier to reconstruct, has not yet shown a clear experimental signal; theoretically, higher-order corrections significantly suppress the LO results, consistent with observations. In contrast, $e^+e^- \to J/\psi+\eta_c$ has been measured to be 3–5 times larger than the LO prediction, with sizable contributions from NLO, next-to-next-to-leading order (NNLO), $\alpha_s v^2$ corrections, and QED diagrams. It should be noted that CO processes can also contribute to exclusive double-charmonium production \cite{Braaten:2002fi}. For instance, two gluons emitted during the transition of one CO $c\bar{c}$ pair to a CS state (which can form a charmonium) may be absorbed by the other CO $c\bar{c}$ pair. This mechanism leads to $v^{4}$ suppression, \footnote{$v$ denotes the relative velocity of the constituent $c\bar{c}$ pair.} which, combined with the suppression required by helicity selection rules, makes CO contributions generally negligible compared with CS processes.

Given the unsettled status of NRQCD factorization in inclusive $J/\psi$ production via $e^+e^-$ annihilation, and the mild role of CO contributions in exclusive double-charmonium channels, it remains of interest to further examine the CO mechanism in the low-energy regime. To this end, this work presents a systematic NLO analysis of inclusive double-charmonium production in $e^+e^-$ annihilation. Since double-$J/\psi$ production has not yet been observed experimentally, we concentrate on the inclusive process $e^+e^- \to J/\psi+\eta_c+X$. Theoretically, this channel receives a $^3S_1^{[8]}$ single-gluon fragmentation (SGF) contribution, which may introduce sizable CO effects and thereby modify the energy dependence of the total cross section. On the experimental side, the Belle Collaboration \cite{Belle:2023gln} has recently demonstrated the feasibility of reconstructing both $J/\psi \to \mu^+\mu^-$ and $\eta_c$ via six charged decay modes in the search for $Y_{cc}$ states, thereby establishing a solid foundation for observing inclusive $J/\psi+\eta_c$ production.

In the literature, inclusive $J/\psi+\eta_c$ production has been studied in both hadroproduction and $e^+e^-$ annihilation. In hadroproduction, ref.~\cite{Lansberg:2013qka} first investigated this channel and found sizable differential cross sections in the intermediate-to-high $p_T$ region, making it accessible to both theory and experiment. For $e^+e^-$ collisions, the first LO study was performed in ref.~\cite{Braaten:2002fi}; however, that work did not include all relevant CO contributions. As we will show later, the higher-order corrections omitted there can yield numerically important effects. The present work is therefore devoted to filling this theoretical gap and providing new insights into the heavy-quarkonium production mechanism.

The paper is structured as follows: Section II details the calculation formalism. Section III presents phenomenological results and analysis. Section IV provides a concise summary.

\section{Calculation formalism}\label{cal}
Within the NRQCD framework \cite{Bodwin:1994jh,Petrelli:1997ge}, the cross section of $e^{+}e^{-} \to J/\psi+\eta_c+X$ can be factorized as
\begin{eqnarray}
\sigma_{e^{+}e^{-} \to J/\psi+\eta_c+X}=\hat{\sigma}_{e^{+}e^{-} \to c\bar{c}[n_1]+c\bar{c}[n_2]+X} \times \langle \mathcal{O}^{J/\psi}(n_1)\rangle \times \langle \mathcal{O}^{\eta_c}(n_2)\rangle,\label{eq3}
\end{eqnarray}
where $\hat{\sigma}_{e^{+}e^{-} \to c\bar{c}[n_1]+c\bar{c}[n_2]+X}$ are the perturbatively calculable short-distance coefficients (SDCs) describing the production of the intermediate $c\bar{c}[n_1]$ and $c\bar{c}[n_2]$ states. For the $J/\psi$ ($^3S_1$) meson, the CO states that first appear in the $v$ expansion are $^1S_0^{[8]}$, $^3S_1^{[8]}$, and $^3P_J^{[8]}$, in addition to the leading $v$ contribution from the CS transition ${}^3S_1^{[1]}$. For the $\eta_c$ ($^1S_0$) meson, the corresponding relevant CS and CO states are $^1S_0^{[1]}$, $^1S_0^{[8]}$, $^3S_1^{[8]}$, and $^1P_1^{[8]}$, respectively. The $\langle \mathcal{O}^{J/\psi}(n_1)\rangle$ and $\langle \mathcal{O}^{\eta_c}(n_2)\rangle$ represent the universal nonperturbative LDMEs.

In calculating $\hat{\sigma}_{e^{+}e^{-} \to c\bar{c}[n_1]+c\bar{c}[n_2]+X}$, the contributing amplitudes are classified into three categories according to the color structure of the intermediate states: CS+CS, CS+CO, and CO+CO processes. For the first two categories, we include all relevant contributions, given the relatively large size of the CS LDMEs. For the CO+CO set, the contributions are suppressed by a factor of $v^8$ due to the two CO LDMEs involved. However, the channels involving the $^3S_1^{[8]}$ SGF topology are expected to significantly compensate for this suppression, thereby yielding sizable contributions. Therefore, for the CO+CO channels, we retain only those involving the SGF topology. The detailed treatment is as follows:
\begin{eqnarray}
\textrm{CS+CS:}&~&e^{+}e^{-} \to c\bar{c}[^3S_1^{[1]}]+c\bar{c}[^1S_0^{[1]}]~(\textrm{NLO}); \nonumber \\
\textrm{CS+CO:}&~&e^{+}e^{-} \to c\bar{c}[^3S_1^{[1]}]+c\bar{c}[^3S_1^{[8]},^1P_1^{[8]}]+g, \nonumber \\
&~&e^{+}e^{-} \to c\bar{c}[^1S_0^{[8]},^3P_J^{[8]}]+c\bar{c}[^1S_0^{[1]}]+g; \nonumber \\
\textrm{CO+CO:}&~&e^{+}e^{-} \to c\bar{c}[^1S_0^{[8]}]+c\bar{c}[^3S_1^{[8]}]~(\textrm{NLO}), \nonumber \\
&~&e^{+}e^{-} \to c\bar{c}[^3S_1^{[8]}]+c\bar{c}[^1S_0^{[8]}]~(\textrm{NLO}), \nonumber \\
&~&e^{+}e^{-} \to c\bar{c}[^3S_1^{[8]}]+c\bar{c}[^3S_1^{[8]},^1P_1^{[8]}]+g, \nonumber \\
&~&e^{+}e^{-} \to c\bar{c}[^3P_J^{[8]}]+c\bar{c}[^1S_0^{[8]}]+g, \nonumber \\
&~&e^{+}e^{-} \to c\bar{c}[^3P_J^{[8]}]+c\bar{c}[^3S_1^{[8]}]~(\textrm{NLO}).
\label{channels}
\end{eqnarray}

It is worth emphasizing that, in addition to the processes explicitly involving the $^{3}S_1^{[8]}$ state, the CO+CO category also includes the process $e^{+}e^{-} \to c\bar{c}[^{3}P_J^{[8]}]+c\bar{c}[^{1}S_0^{[8]}]+g$. In this channel, the emission of a soft gluon from the final-state $^{3}P_J^{[8]}$ induces soft infrared (IR) divergences, which can be canceled through the renormalization of the LDME $\langle \mathcal{O}^{^{3}P_J^{[8]}}(^{3}S_1^{[8]})\rangle$ associated with the LO SDCs of $e^{+}e^{-} \to c\bar{c}[^{3}S_1^{[8]}]+c\bar{c}[^{1}S_0^{[8]}]$. Consequently, the process $e^{+}e^{-} \to c\bar{c}[^{3}P_J^{[8]}]+c\bar{c}[^{1}S_0^{[8]}]+g$ also effectively includes a $^{3}S_1^{[8]}$ SGF contribution. In contrast, for processes not considered in this work, such as $e^{+}e^{-} \to c\bar{c}[^{1}S_0^{[8]}]+c\bar{c}[^{1}P_1^{[8]}]$ and $e^{+}e^{-} \to c\bar{c}[^{3}P_J^{[8]}]+c\bar{c}[^{1}P_1^{[8]}]$, the amplitudes contain neither $g^* \to c\bar{c}[^{3}S_1^{[8]}]$ nor $g^* \to c\bar{c}[^{3}P_J^{[8]}]+g$ structures at LO or NLO. Since their contributions suffer solely from a $v^8$ suppression without any kinematic compensation from the SGF topology, omitting them is not expected to compromise the precision of our calculations.

\begin{figure}[!h]
\begin{center}
\hspace{0cm}\includegraphics[width=0.75\textwidth]{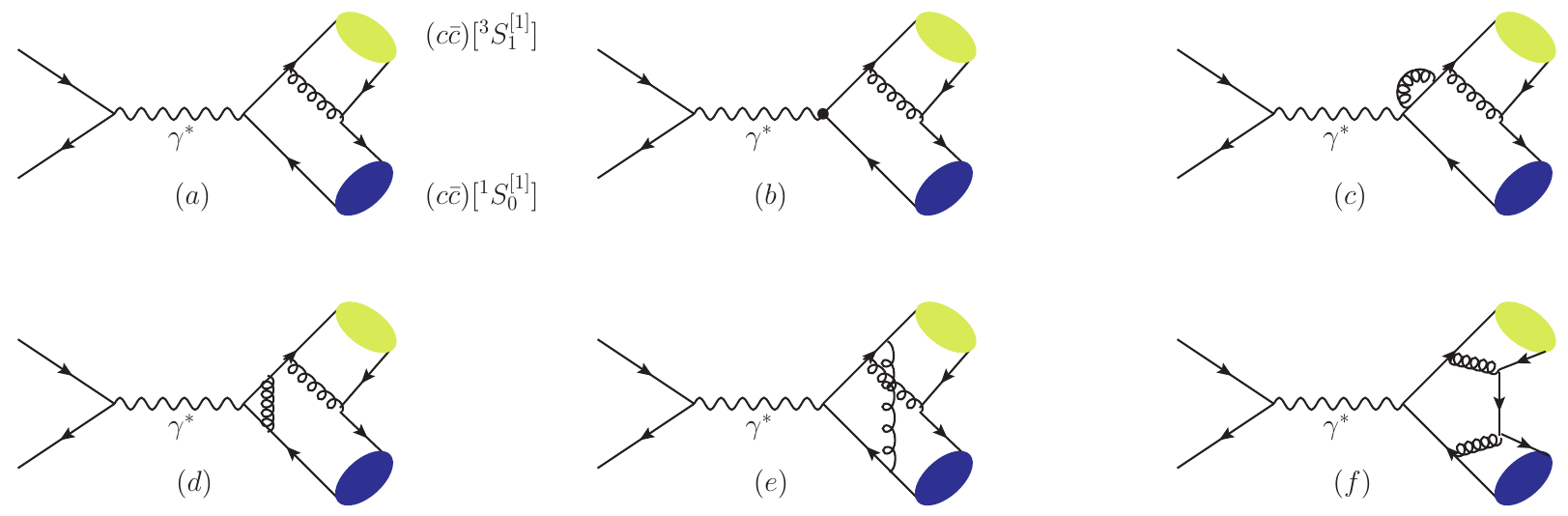}
\caption{\label{CS_QCD}
Representative QCD Feynman diagrams for $e^{+}e^{-} \to c\bar{c}[^3S_1^{[1]}]+c\bar{c}[^1S_0^{[1]}]$.(a)($\alpha\alpha_s$ order) is the QCD tree-level diagram; (b-f)($\alpha\alpha_s^2$ order) are the NLO QCD corrections to (a). The diagram (b) with heavy dot denotes the counter-term diagram.}
\end{center}
\end{figure}

\begin{figure}
\includegraphics[width=0.95\textwidth]{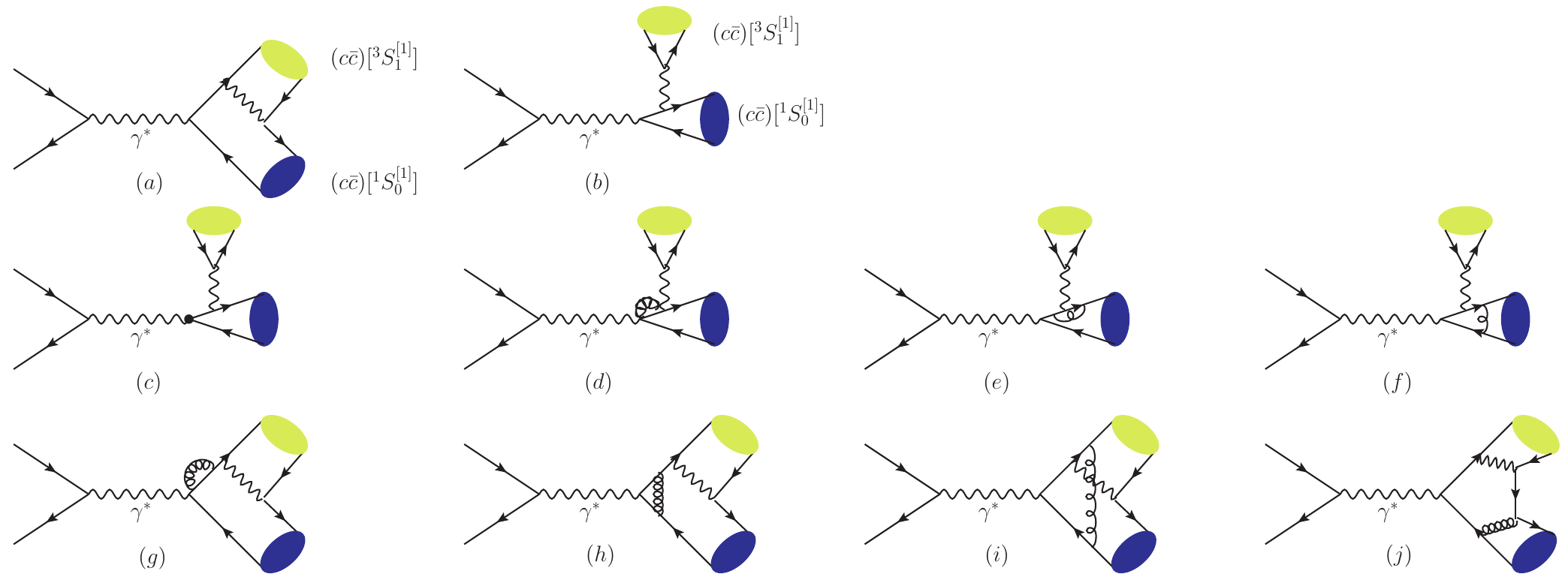}
\caption{\label{CS_QED}
Representative QED Feynman diagrams for $e^{+}e^{-} \to c\bar{c}[^3S_1^{[1]}]+c\bar{c}[^1S_0^{[1]}]$. (a,b)($\alpha^2$ order) are the QED tree-level diagrams; (c-j)($\alpha^2\alpha_s$) are the NLO QCD corrections to (a,b). The diagram (c) with heavy dot denotes the counter-term diagram.}
\end{figure}

\subsection{CS+CS process}

In the calculation of the CS+CS channel, $e^{+}e^{-} \to c\bar{c}[^{3}S_1^{[1]}]+c\bar{c}[^{1}S_0^{[1]}]$, we include both the conventional QCD diagrams (figure \ref{CS_QCD}) and the QED contributions. This is motivated by the fact that the interference between the QCD amplitude and the QED diagrams (figure \ref{CS_QED}) containing the single-photon fragmentation (SPF) contribution $\gamma^* \to c\bar{c}[^{3}S_1^{[1]}]$ has been found to be non-negligible in previous studies \cite{Braaten:2002fi,Liu:2004ga}. In our previous work (ref. \cite{Sun:2018rgx}), we successfully reproduced the pure QCD NLO results ($K$ factor) listed in table 1 of ref. \cite{Gong:2007db} by adopting the same parameter set, and for the first time, we performed the QCD corrections to the QED--QCD interference, although the analysis was limited to $\sqrt{s}=\Upsilon(4S)$. In this work, we extend the CS results to a wider energy range, explicitly illustrating that the contribution of this interference term becomes increasingly significant as the collision energy increases. Following ref. \cite{Sun:2018rgx}, we briefly outline our strategy for computing the combined QCD+QED contributions below.

First, we further express the SDCs of $e^{+}e^{-} \to c\bar{c}[^3S_1^{[1]}]+c\bar{c}[^1S_0^{[1]}]$ as
\begin{eqnarray}
\hat{\sigma}_{e^{+}e^{-} \to c\bar{c}[^3S_1^{[1]}]+c\bar{c}[^1S_0^{[1]}]}=\int |\mathcal{M}|^2 d\Pi_2,
\end{eqnarray}
where $|\mathcal{M}|^2$ is the squared matrix elements and $d\Pi_2$ is the standard two-body phase space. Up to the $\alpha^3$ order and NLO accuracy in $\alpha_s$, the $|\mathcal{M}|^2$ can be written as
\begin{eqnarray}
&&\big| \left(\mathcal{M}_{\alpha\alpha_s}+\mathcal{M}_{\alpha\alpha_s^2}\right) + \left(\mathcal{M}_{\alpha^2}+\mathcal{M}_{\alpha^2\alpha_s}\right)\big|^2 \nonumber \\
&=&\big|\mathcal{M}_{\alpha\alpha_s}\big|^2+2\textrm{Re}\left(\mathcal{M}^{*}_{\alpha\alpha_s}\mathcal{M}_{\alpha\alpha_s^2}\right)+2\textrm{Re}\left(\mathcal{M}^{*}_{\alpha\alpha_s}\mathcal{M}_{\alpha^2}\right) \nonumber \\
&~&+2\textrm{Re}\bigg[\left(\mathcal{M}^{*}_{\alpha\alpha_s}\mathcal{M}_{\alpha^2\alpha_s}\right)+\left(\mathcal{M}_{\alpha\alpha_s^2}\mathcal{M}^{*}_{\alpha^2}\right)\bigg]+\cdots.\label{squared matrix elements}
\end{eqnarray}
The Feynman diagrams contributing to the amplitudes $\mathcal{M}_{\alpha\alpha_s}$, $\mathcal{M}_{\alpha\alpha_s^2}$, $\mathcal{M}_{\alpha^2}$, and $\mathcal{M}_{\alpha^2\alpha_s}$ are grouped according to their coupling orders and are representatively shown in figures \ref{CS_QCD} and~\ref{CS_QED}. Specifically, figure \ref{CS_QCD}(a), of order $\alpha\alpha_s$, depicts the QCD tree-level diagrams, while figures \ref{CS_QCD}(b--f), of order $\alpha\alpha_s^2$, represent their NLO QCD corrections. Meanwhile, figures \ref{CS_QED}(a--b), of order $\alpha^2$, correspond to the QED tree-level contributions, and figures \ref{CS_QED}(c--j), of order $\alpha^2\alpha_s$, illustrate the higher-order QCD corrections to the QED diagrams. 

It is worth noting that, in the calculation of $\mathcal{M}_{\alpha^2\alpha_s}$, we include only the QCD corrections to the QED diagrams, as shown in figure \ref{CS_QED}; the QED corrections to figure \ref{CS_QCD}(a) are omitted as an approximation. This approximation is justified as follows. At the tree level, the QED--QCD interference enhances the pure QCD result by about $20\%$ \cite{Braaten:2002fi,Liu:2004ga}, arising mainly from the SPF structure in the QED diagrams. The kinematic enhancement due to SPF can partially offset the $\alpha/\alpha_s$ suppression of the QED diagrams relative to the QCD ones, thereby producing a sizable contribution. At NLO, however, while the QCD corrections to the QED diagrams preserve the SPF topology, the QED corrections to the QCD tree diagrams do not. Therefore, without this kinematic enhancement, the latter are expected to be considerably smaller than the former. A definitive assessment of this approximation, however, requires a complete QED NLO calculation.

According to equation (\ref{squared matrix elements}), we decompose the SDCs into four parts,
\begin{eqnarray}
\hat{\sigma}=\hat{\sigma}_{\alpha^2}^{(0)}+\hat{\sigma}_{\alpha^2}^{(1)}+\hat{\sigma}_{\alpha^3}^{(0)}+\hat{\sigma}_{\alpha^3}^{(1)},
\end{eqnarray}
where
\begin{eqnarray}
\hat{\sigma}_{\alpha^2}^{(0)} &= & \int \big|\mathcal{M}_{\alpha\alpha_s} \big|^2d\Pi_{2}, \nonumber \\
\hat{\sigma}_{\alpha^2}^{(1)} &= & \int 2\textrm{Re}\big(\mathcal{M}^{*}_{\alpha\alpha_s} \mathcal{M}_{\alpha\alpha_s^2}\big)d\Pi_{2}, \nonumber \\
\hat{\sigma}_{\alpha^3}^{(0)} &= & \int 2\textrm{Re}\big(\mathcal{M}^{*}_{\alpha\alpha_s} \mathcal{M}_{\alpha^2}\big)d\Pi_{2}, \nonumber \\
\hat{\sigma}_{\alpha^3}^{(1)} &= & \int 2\textrm{Re}\bigg[\left(\mathcal{M}^{*}_{\alpha\alpha_s}\mathcal{M}_{\alpha^2\alpha_s}\right)+\left(\mathcal{M}_{\alpha\alpha_s^2}\mathcal{M}^{*}_{\alpha^2}\right)\bigg]d\Pi_{2}.
\label{Eq. QED+QCD}
\end{eqnarray}

We employ dimensional regularization with $D = 4 - 2\epsilon$ to regularize both ultraviolet (UV) and IR divergences. The renormalization constants for the charm-quark mass ($Z_m$) and the heavy-quark field ($Z_2$) are defined in the on-mass-shell (OS) scheme; while the minimal-subtraction ($\overline{\mathrm{MS}}$) scheme is adopted for the QCD gauge coupling ($Z_g$) and the gluon field ($Z_3$). The corresponding renormalization constants are given as \cite{Zhang:2005cha}
\begin{eqnarray}
\delta Z_{m}^{OS}&=& -3 C_{F} \frac{\alpha_s}{4\pi}\left[\frac{1}{\epsilon_{\textrm{UV}}}-\gamma_{E}+\textrm{ln}\frac{4 \pi \mu_r^2}{m_c^2}+\frac{4}{3}\right], \nonumber \\
\delta Z_{2}^{OS}&=& - C_{F} \frac{\alpha_s}{4\pi}\left[\frac{1}{\epsilon_{\textrm{UV}}}+\frac{2}{\epsilon_{\textrm{IR}}}-3 \gamma_{E}+3 \textrm{ln}\frac{4 \pi \mu_r^2}{m_c^2}+4\right], \nonumber \\
\delta Z_{3}^{\overline{MS}}&=& \frac{\alpha_s}{4\pi}(\beta_{0}-2 C_{A})\left[\frac{1}{\epsilon_{\textrm{UV}}}-\gamma_E+\textrm{ln}(4\pi)\right], \nonumber \\
\delta Z_{g}^{\overline{MS}}&=& -\frac{\beta_{0}}{2}\frac{\alpha_s}{4\pi}\left[\frac{1} {\epsilon_{\textrm{UV}}}-\gamma_{E}+\textrm{ln}(4\pi)\right], \label{CT}
\end{eqnarray}
where $\gamma_E$ is the Euler constant, and $\beta_0 = \frac{11}{3}C_A - \frac{4}{3}T_F n_f$ is the one-loop coefficient of the QCD $\beta$-function. Here, $n_f = n_L + n_H$ denotes the number of active quark flavors, with $n_L$ and $n_H$ representing the numbers of light- and heavy-quark flavors, respectively. In the energy region around the $\Upsilon(4S)$ in this work, we take $n_L = 3$ and $n_H = 1$. The color factors in ${\rm SU}(3)$ are $T_F = \frac{1}{2}$, $C_F = \frac{4}{3}$, and $C_A = 3$.

\subsection{CS+CO process}

\begin{figure}[!h]
\begin{center}
\hspace{0cm}\includegraphics[width=0.65\textwidth]{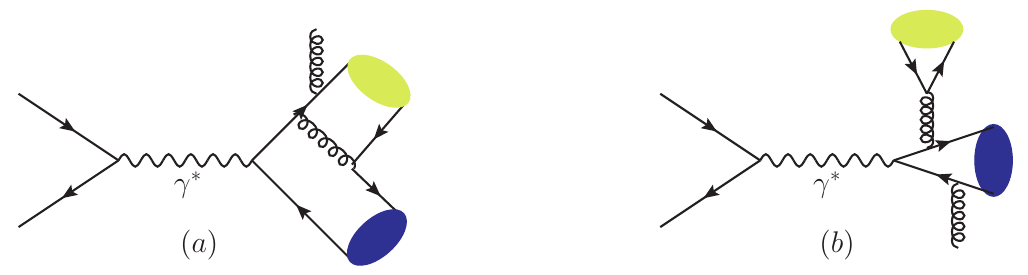}
\caption{\label{CS+CO}
Representative Feynman diagrams for the CS+CO processes. The left panel displays diagrams for both $e^{+}e^{-} \to c\bar{c}[^{3}S_1^{[1]}]+c\bar{c}[^{3}S_1^{[8]},^{1}P_1^{[8]}]+g$ and $e^{+}e^{-} \to c\bar{c}[^{1}S_0^{[8]},^{3}P_J^{[8]}]+c\bar{c}[^{1}S_0^{[1]}]+g$. The right panel exclusively shows diagrams for the channel involving the $^{3}S_1^{[8]}$ state.}
\end{center}
\end{figure}

During the CS+CO processes in equation (\ref{channels}), the channels $e^{+}e^{-} \to c\bar{c}[^{3}S_1^{[1]}]+c\bar{c}[^{3}S_1^{[8]}]+g$ and $e^{+}e^{-} \to c\bar{c}[^{1}S_0^{[8]}]+c\bar{c}[^{1}S_0^{[1]}]+g$ are free of IR divergences and can be computed directly by evaluating the corresponding Feynman diagrams (figure \ref{CS+CO}). In contrast, the channels $e^{+}e^{-} \to c\bar{c}[^{3}S_1^{[1]}]+c\bar{c}[^{1}P_1^{[8]}]+g$ and $e^{+}e^{-} \to c\bar{c}[^{3}P_J^{[8]}]+c\bar{c}[^{1}S_0^{[1]}]+g$ involve soft gluons coupling to the P-wave states, which induce soft IR singularities that must be absorbed into the renormalization of the corresponding LDMEs. In the following, we take the process $e^{+}e^{-} \to c\bar{c}[^{3}P_J^{[8]}]+c\bar{c}[^{1}S_0^{[1]}]+g$ as an example to illustrate the procedure.

To derive the finite SDCs $\hat{\sigma}_{e^{+}e^{-} \to c\bar{c}[^{3}P_J^{[8]}]+c\bar{c}[^{1}S_0^{[1]}]+g}$ (denoted as $\hat{\sigma}(^3P_{J}^{[8]})$ for convenience), we begin by decomposing the corresponding cross section $\sigma(^3P_{J}^{[8]})$ into two parts,\footnote{Owing to the absence of the process $\sigma_{e^{+}e^{-} \to c\bar{c}[^{3}S_1^{[8]}]+c\bar{c}[^{1}S_0^{[1]}]}$, only $\langle \mathcal{O}^{^3P_J^{[8]}}(^3S_1^{[1]}) \rangle$ requires renormalization.}
\begin{eqnarray}
\sigma(^3P_{J}^{[8]})=\hat{\sigma}_{^3P_{J}^{[8]}}\langle \mathcal O ^{J/\psi}(^3P_{J}^{[8]})\rangle+\hat{\sigma}^{\textrm{LO}}_{^3S_{1}^{[1]}} \langle \mathcal O ^{^3P_{J}^{[8]}}(^3S_{1}^{[1]})\rangle^{\textrm{renorm}}\langle \mathcal O ^{J/\psi}(^3P_{J}^{[8]})\rangle,
\end{eqnarray}
then one can obtain
\begin{eqnarray}
\hat{\sigma}_{^3P_{J}^{[8]}}\langle \mathcal O ^{J/\psi}(^3P_{J}^{[8]})\rangle
&=&\sigma(^3P_{J}^{[8]})-\hat{\sigma}^{\textrm{LO}}_{^3S_{1}^{[1]}} \langle \mathcal O ^{^3P_{J}^{[8]}}(^3S_{1}^{[1]})\rangle^{\textrm{renorm}} \langle \mathcal O ^{J/\psi}(^3P_{J}^{[8]})\rangle\nonumber \\
&=&\left[\left(\hat{\sigma}_\textrm{S}+\hat{\sigma}_{\textrm{H}}\right)\big|_{^3P_{J}^{[8]}}-\hat{\sigma}^{\textrm{LO}}_{^3S_{1}^{[1]}} \langle \mathcal O ^{^3P_{J}^{[8]}}(^3S_{1}^{[1]})\rangle^{\textrm{renorm}}\right] \langle \mathcal O ^{J/\psi}(^3P_{J}^{[8]})\rangle. \nonumber
\\
\label{3pj8eq1}
\end{eqnarray}

The small cutoff parameter $\delta_s$ adopted in the two-cutoff slicing method \cite{Harris:2001sx} is introduced to separate the phase space into a soft region ($\hat{\sigma}_{\text{S}}$) and a hard region ($\hat{\sigma}_{\text{H}}$). $\hat{\sigma}_{\text{H}}$ is IR-finite and is evaluated numerically using standard Monte Carlo integration techniques. Following the strategy adopted in ref.~\cite{Zhang:2014coi}, the soft part $\hat{\sigma}_{\text{S}}$ for the process $e^{+}e^{-} \to c\bar{c}[^{3}P_J^{[8]}]+c\bar{c}[^{1}S_0^{[1]}]+g$ can be expressed as analytically in terms of the LO SDCs $\hat{\sigma}^{\mathrm{LO}}_{^{3}S_1^{[1]}}$ for the partonic process $e^{+}e^{-} \to c\bar{c}[^{3}S_1^{[1]}]+c\bar{c}[^{1}S_0^{[1]}]$. Subsequently, the soft singularities in $\hat{\sigma}_{\mathrm{S}}$ cancel with the IR divergences in $\langle \mathcal{O} ^{^3P^{[8]}_J}(^3S^{[1]}_1) \rangle^\mathrm{renorm}$, yielding a finite result for $\hat{\sigma}_{^3P^{[8]}_J}$,

\begin{eqnarray}
\hat{\sigma}_{^3P_{J}^{[8]}}=\hat{\sigma}_{\textrm{H}}+\hat{\sigma}_{\textrm{S}}^{*},
\label{3pj8+1s01 SDCs}
\end{eqnarray}
with
\begin{eqnarray}
\hat{\sigma}_{\textrm{S}}^{*}=-\frac{4\alpha_s}{3 \pi m^2_c}u_{\epsilon}\hat{\sigma}^{\textrm{LO}}_{^3S_{1}^{[1]}}.
\end{eqnarray}
The analytical expression for $u_{\epsilon}$ is given in ref. \cite{Zhang:2014coi}; specifically,
\begin{eqnarray}
u_{\epsilon}=\frac{p_0}{|\bf{p}|}\textrm{ln}\left(\frac{p_0+|\bf{p}|}{p_0-|\bf{p}|}\right)+\textrm{ln}\left(\frac{\mu_{\Lambda}^2}{s\delta_s^2}\right)-2+2\textrm{ln}(2),
\label{u_epsilon}
\end{eqnarray}
where $p_0$ and $\mathbf{p}$ denote the energy and three-momentum of the $J/\psi$, respectively. $\mu_\Lambda$ denotes the scale arising from the renormalization of the LDME and is set to $\mu_\Lambda = m_c$ \cite{Zhang:2014coi}.

For the channel $e^{+}e^{-} \to c\bar{c}[^{3}S_1^{[1]}]+c\bar{c}[^{1}P_1^{[8]}]+g$, the finite SDCs can be obtained through the renormalization of the LDME $\langle \mathcal{O}^{^{1}P_{1}^{[8]}}(^{1}S_{0}^{[1]})\rangle$, following the formulas given in equations (\ref{3pj8+1s01 SDCs}-\ref{u_epsilon}).

\subsection{CO+CO processes}

For the CO+CO processes, we compute the QCD corrections to the channels $e^{+}e^{-} \to c\bar{c}[^{1}S_0^{[8]},^{3}P_J^{[8]}]+c\bar{c}[^{3}S_1^{[8]}]$. Up to NLO accuracy in $\alpha_s$, the SDCs can be expressed as
\begin{eqnarray}
\hat{\sigma}=\hat{\sigma}_{\textrm{LO}}+\hat{\sigma}_{\textrm{V}}+\hat{\sigma}_{\textrm{R}},
\end{eqnarray}
where $\hat{\sigma}_{\text{V}}$ represents the virtual corrections, including counterterms and one-loop corrections. The corresponding diagrams are obtained from figure \ref{CS_QED} by replacing the photon propagators with gluon ones. The renormalization scheme and the corresponding constants are specified in equation (\ref{CT}). $\hat{\sigma}_{\text{R}}$ denotes the real corrections, corresponding to the process $e^{+}e^{-} \to c\bar{c}[^{1}S_0^{[8]},^{3}P_J^{[8]}]+c\bar{c}[^{3}S_1^{[8]}]+g$, with the diagrams shown in figure \ref{CS+CO}. Using the two-cutoff slicing method \cite{Harris:2001sx}, $\hat{\sigma}_{\text{R}}$ is decomposed into soft ($\hat{\sigma}_{\text{S}}$) and hard ($\hat{\sigma}_{\text{H}}$) terms. Note that for the channel $e^{+}e^{-} \to c\bar{c}[^{3}P_J^{[8]}]+c\bar{c}[^{3}S_1^{[8]}]+g$, since the process $e^{+}e^{-} \to c\bar{c}[^{3}S_1^{[1]},^{3}S_1^{[8]}]+c\bar{c}[^{3}S_1^{[8]}]$ is absent, the cancellation of IR singularities arising from soft gluons attaching to the final-state $^{3}P_J^{[8]}$ does not require the renormalization of the LDME $\langle \mathcal{O}^{^{3}P_{J}^{[8]}}(^{3}S_{1}^{[1]},^{3}S_{1}^{[8]}) \rangle$. Consequently, the combination of $\hat{\sigma}_{\text{Virtual}}$ and $\hat{\sigma}_{\text{Real}}$ yields a divergence-free result.

As for the channel $e^{+}e^{-} \to c\bar{c}[^{3}S_1^{[8]}]+c\bar{c}[^{3}S_1^{[8]}]+g$, no soft-gluon divergences appear, and the amplitudes can be evaluated directly from the relevant diagrams. However, for the channels $e^{+}e^{-} \to c\bar{c}[^{3}S_1^{[8]}]+c\bar{c}[^{1}P_1^{[8]}]+g$ and $e^{+}e^{-} \to c\bar{c}[^{3}P_J^{[8]}]+c\bar{c}[^{1}S_0^{[8]}]+g$, soft IR divergences arise again due to the possible emission of soft gluons from the P-wave states. These divergences are absorbed through the renormalization of the corresponding LDMEs, $\langle \mathcal{O}^{^{1}P_{1}^{[8]}}(^{1}S_{0}^{[8]})\rangle$ and $\langle \mathcal{O}^{^{3}P_{J}^{[8]}}(^{3}S_{1}^{[8]})\rangle$, respectively. By analogy with equations (\ref{3pj8+1s01 SDCs}-\ref{u_epsilon}) and following the strategy in ref. \cite{Zhang:2014coi}, the finite SDCs for $e^{+}e^{-} \to c\bar{c}[^{3}S_1^{[8]}]+c\bar{c}[^{1}P_1^{[8]}]+g$ (denoted as $\hat{\sigma}_{^{1}P_1^{[8]}}$) can be written as
\begin{eqnarray}
\hat{\sigma}_{^{1}P_1^{[8]}}=\hat{\sigma}_{\textrm{H}}+\hat{\sigma}^{*}_{\textrm{S}},
\label{3s18+1p18 SDCs}
\end{eqnarray}
with
\begin{eqnarray}
\hat{\sigma}^{*}_{\textrm{S}}=-\frac{\alpha_s}{3 \pi m^2_c}\frac{N_c^2-4}{N_c}u_{\epsilon}\hat{\sigma}^{\textrm{LO}}_{^1S_{0}^{[8]}}.
\end{eqnarray}
Here $\hat{\sigma}^{\mathrm{LO}}_{^{1}S_0^{[8]}}$ denotes the LO SDCs for the partonic process $e^{+}e^{-} \to c\bar{c}[^{3}S_1^{[8]}]+c\bar{c}[^{1}S_0^{[8]}]$. For the process $e^{+}e^{-} \to c\bar{c}[^{3}P_J^{[8]}]+c\bar{c}[^{1}S_0^{[8]}]+g$, the corresponding finite SDCs can be obtained straightforwardly by applying equation (\ref{3s18+1p18 SDCs}).

In our calculations of the CS+CS, CS+CO, and CO+CO channels, we employ a Mathematica-Fortran package that integrates FeynArts~\cite{Hahn:2000kx}, FeynCalc~\cite{Mertig:1990an}, FIRE~\cite{Smirnov:2008iw}, and Apart~\cite{Feng:2012iq}. This package has been applied to various heavy-quarkonium processes up to NLO accuracy in $\alpha_s$~\cite{Sun:2018rgx,Sun:2021tma,Li:2023tzx,Sun:2017wxk,Sun:2021avu,Jiang:2025miz}. As a cross-check, we also compute all relevant processes using the independent Feynman Diagram Calculation (FDC) package~\cite{Wang:2004du}, and find consistent numerical results.

\section{Phenomenological Results}

\subsection{Input Parameters}
In the calculations, we take $M_{J/\psi}=M_{\eta_c}=2m_c$ with $m_c=1.5~\text{GeV}$, and fix the fine-structure constant at $\alpha=1/137$ \cite{Liu:2002wq}. The strong coupling $\alpha_s$ is evolved using the two-loop running with $\Lambda^{(4)}_{\overline{\textrm{MS}}}=338$ MeV, with which one can obtain $\alpha_s\big|_{\mu_r=3~\textrm{GeV}}=0.259$ \cite{Gong:2007db}. The cutoff parameter $\delta_s$ is set to $0.001$, and the stability of the results with respect to its variation will be demonstrated later. Given the large integrated luminosity accumulated by the Belle Collaboration at $\sqrt{s}=10.52$~GeV, $\Upsilon(4S)$ (10.58 GeV), and $\Upsilon(5S)$ (10.89 GeV), we present our results at these three energy points. Moreover, to more clearly illustrate the role of the CO mechanism, we extend the calculations up to $20$~GeV, which covers the threshold region relevant for $\Upsilon(4S)$ pair production.
\begin{table*}[htb]
\begin{center}
\caption{The values of the $J/\psi$-related LDMEs.}
\label{tab: LDMEs}
\begin{tabular}{lccccccc}
\hline\hline
$~~~$ & $\langle \mathcal O ^{J/\psi}(^3S_1^{[1]}) \rangle$ & $\langle \mathcal O ^{J/\psi}(^1S_0^{[8]}) \rangle$ & $\langle \mathcal O ^{J/\psi}(^3S_1^{[8]}) \rangle$ & $\langle \mathcal O ^{J/\psi}(^3P_J^{[8]}) \rangle/m_c^2$\\
$~~~$ & $\textrm{GeV}^3$ & $10^{-2}\textrm{GeV}^3$ & $10^{-2}\textrm{GeV}^3$ & $10^{-2}\textrm{GeV}^3$\\ \hline
set 1 \cite{Butenschoen:2010rq} & $1.32$ & $4.50 \pm 0.72$ & $0.312 \pm 0.93$ & $-0.538 \pm 0.156$\\ \hline
set 2 \cite{Chao:2012iv} & $1.16$ & $8.9 \pm 0.98$ & $0.30 \pm 0.12$ & $0.56 \pm 0.21$\\ \hline
set 3 \cite{Bodwin:2014gia} & $1.16$ & $9.9 \pm 2.2$ & $1.1 \pm 1.0$ & $0.489 \pm 0.444$\\ \hline
set 4 \cite{Brambilla:2024iqg} & $1.16$ & $0.068 \pm 0.2489$ & $1.050 \pm 0.121$ & $1.879 \pm 0.261$\\ \hline \hline
\end{tabular}
\end{center}
\end{table*}
As the core input parameters of the NRQCD framework, the LDMEs play a crucial role in theoretical predictions. In this work, we adopt four representative sets of $J/\psi$-related LDMEs from the current literature, taken from refs. \cite{Butenschoen:2010rq,Chao:2012iv,Bodwin:2014gia,Brambilla:2024iqg}, and label them as set 1 through set 4. All four sets provide reasonably good descriptions of the $J/\psi$ hadroproduction data. The numerical values of these LDMEs are listed in table \ref{tab: LDMEs}.

For the $\eta_c$-related LDMEs, we derive them directly from the corresponding $J/\psi$ matrix elements using heavy-quark spin symmetry (HQSS) \cite{Bodwin:1994jh}, with the explicit relations given as follows:

\begin{eqnarray}
\langle \mathcal O ^{\eta_c}(^1S_0^{[1]}) \rangle & \approx &\frac{1}{3} \langle \mathcal O ^{J/\psi}(^3S_1^{[1]}) \rangle, \nonumber \\
\langle \mathcal O ^{\eta_c}(^3S_1^{[8]}) \rangle &\approx & \langle \mathcal O ^{J/\psi}(^1S_0^{[8]}) \rangle, \nonumber \\
\langle \mathcal O ^{\eta_c}(^1S_0^{[8]}) \rangle & \approx &\frac{1}{3} \langle \mathcal O ^{J/\psi}(^3S_1^{[8]}) \rangle, \nonumber \\
\langle \mathcal O ^{\eta_c}(^1P_1^{[8]}) \rangle &\approx&3 \langle \mathcal O ^{J/\psi}(^3P_0^{[8]}). \label{HQSS relations}
\end{eqnarray}

Note that, according to ref.~\cite{Bodwin:1994jh}, HQSS only provides approximate relations between NRQCD matrix elements, and the leading violations of HQSS arise from spin-flip effects, which are of relative order $v^2$. Consequently, the equalities in equation (\ref{HQSS relations}) that follow from HQSS hold only up to relative order $v^2 \sim 0.3$.
\subsection{SDCs}

\begin{figure}[!h]
\begin{center}
\hspace{0cm}\includegraphics[width=0.47\textwidth]{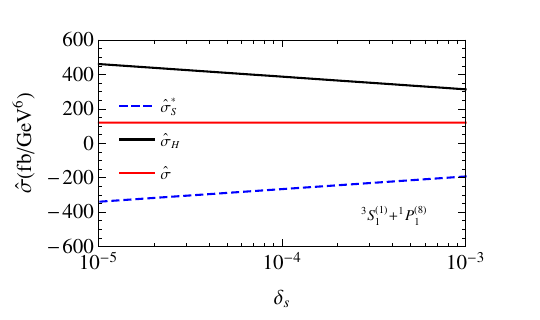}
\hspace{0cm}\includegraphics[width=0.47\textwidth]{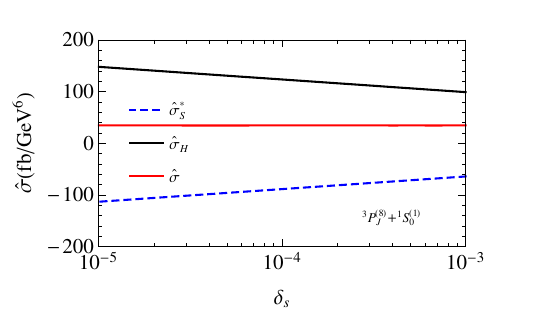}
\hspace{0cm}\includegraphics[width=0.47\textwidth]{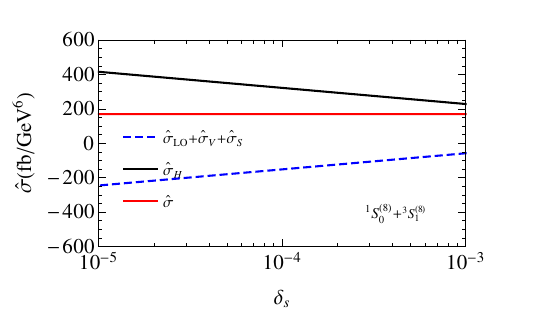}
\hspace{0cm}\includegraphics[width=0.47\textwidth]{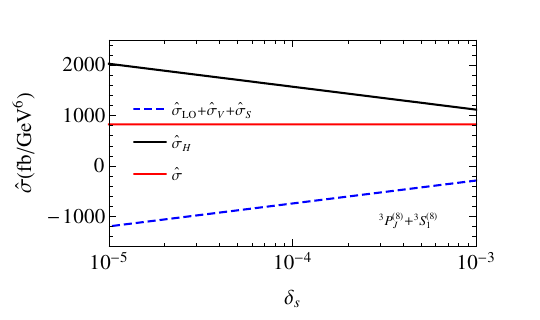}
\hspace{0cm}\includegraphics[width=0.47\textwidth]{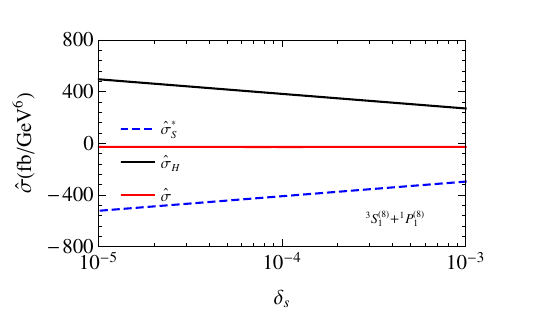}
\hspace{0cm}\includegraphics[width=0.47\textwidth]{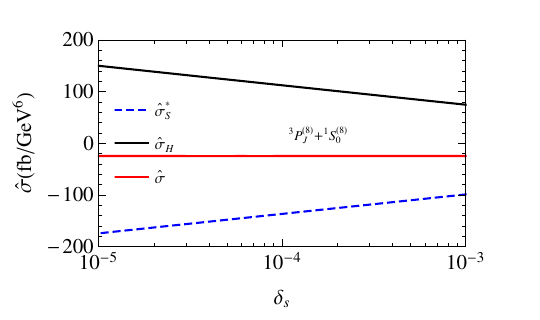}
\caption{\label{deltas running}
Dependence of the results on the soft cutoff parameter $\delta_s$ at $\sqrt{s}=10.58$~GeV. The renormalization scale is fixed at $\mu_r=2m_c$, with $m_c=1.5$~GeV.}
\end{center}
\end{figure}

Before presenting the SDCs for the individual production channels, we first demonstrate the stability of the relevant processes with respect to the parameter $\delta_s$. As shown in figure \ref{deltas running}, for the six processes considered in this work that employ the phase-space slicing method, the calculated results exhibit clear stability against variations of $\delta_s$.

\begin{figure}[!h]
\begin{center}
\hspace{0cm}\includegraphics[width=0.47\textwidth]{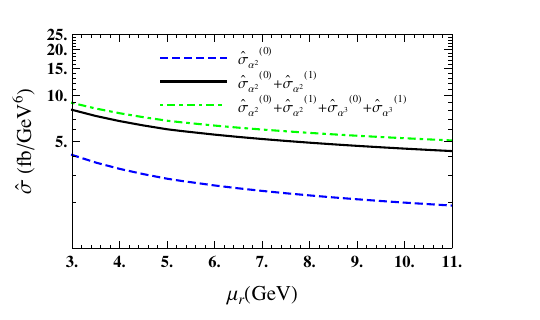}
\hspace{0cm}\includegraphics[width=0.47\textwidth]{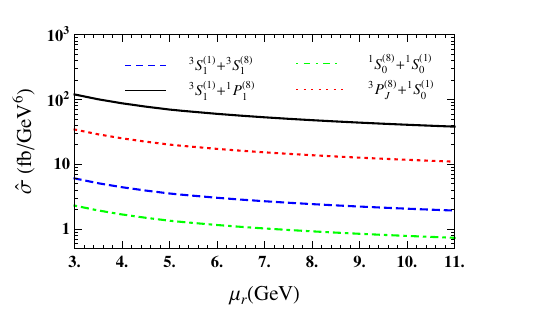}
\hspace{0cm}\includegraphics[width=0.47\textwidth]{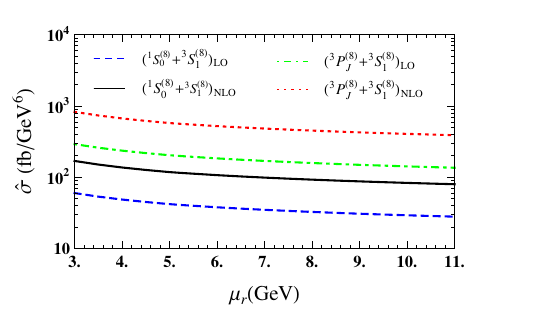}
\hspace{0cm}\includegraphics[width=0.47\textwidth]{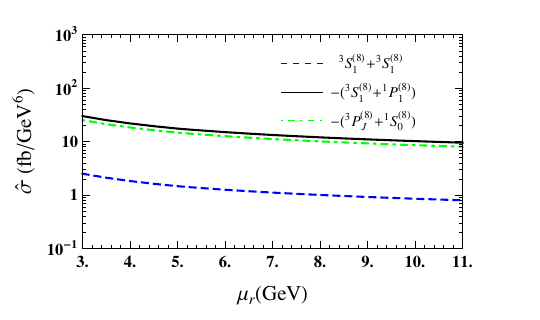}
\caption{\label{SDC}
The SDCs for the CS+CS, CS+CO, and CO+CO processes as a function of the renormalization scale $\mu_r$ at $\sqrt{s}=10.58$~GeV with $m_c=1.5$ GeV. The left and right panels of the first column show the CS+CS and CS+CO results, respectively, while the second column displays the CO+CO results.}
\end{center}
\end{figure}

In figure\ref{SDC}, we present the $\mu_r$-dependence of the SDCs at $\sqrt{s}=10.58$~GeV. As shown in this figure, several key features emerge:
\begin{itemize}
    \item \textbf{CS+CS:} \\
    QCD NLO corrections enhance the LO results by about a factor of two, while the interference between QED and QCD diagrams adds another $10\%$--$20\%$ to the NLO predictions. According to refs.~\cite{Huang:2022dfw,Li:2025mng,Chen:2025qgy}, NNLO corrections can further increase the NLO results by roughly $40\%$. Thus, although the NNLO effects are substantial, the QED contributions remain theoretically important. Notably, as will be shown later, the SPF structure causes the QED contributions to grow rapidly with increasing collision energy, further underscoring their phenomenological relevance.

    \item \textbf{CS+CO:} \\
    A comparison with the CS+CS process shows that the SDCs for the $^{3}S_1^{[1]}+^{1}P_1^{[8]}$ channel are more than one order of magnitude larger than the corresponding $^{3}S_1^{[1]}+^{1}S_0^{[1]}$ values, while those for $^{3}P_J^{[8]}+^{1}S_0^{[1]}$ and $^{3}S_1^{[1]}+^{3}S_1^{[8]}$ are of similar size. In contrast, the coefficients for $^{1}S_0^{[8]}+^{1}S_0^{[1]}$ are significantly suppressed, being several times smaller than the CS+CS results.

    \item \textbf{CO+CO:} \\
    The left panel of the second column reveals that the QCD corrections to the process $e^{+}e^{-} \to c\bar{c}[^{1}S_0^{[8]},^{3}P_J^{[8]}]+c\bar{c}[^{3}S_1^{[8]}]$ are remarkably large, enhancing the corresponding LO results by nearly a factor of three. After including the NLO contributions, the SDCs for these two channels become significantly larger than those of the CS+CS process. Specifically, the SDC for the $^{1}S_0^{[8]}+^{3}S_1^{[8]}$ channel is about 20 times larger than that of the CS+CS process, while the $^{3}P_J^{[8]}+^{3}S_1^{[8]}$ channel exceeds the CS+CS value by nearly two orders of magnitude. For the process $e^{+}e^{-} \to c\bar{c}[^{3}S_1^{[8]}]+c\bar{c}[^{3}S_1^{[8]}]+g$, the SDCs are comparable in magnitude to those of the CS+CS process. In contrast, for the processes $e^{+}e^{-} \to c\bar{c}[^{3}S_1^{[8]}]+c\bar{c}[^{1}P_1^{[8]}]+g$ and $e^{+}e^{-} \to c\bar{c}[^{3}P_J^{[8]}]+c\bar{c}[^{1}S_0^{[8]}]+g$, the finite SDCs contain terms with an opposite sign relative to the large LO SDCs of $e^{+}e^{-} \to c\bar{c}[^{3}S_1^{[8]}]+c\bar{c}[^{1}S_0^{[8]}]$, resulting in negative values with magnitudes approximately three times larger than the CS+CS SDCs.
\end{itemize}

In summary, we have systematically compared the SDCs across different production channels. Among them, the CO+CO channel $e^{+}e^{-} \to c\bar{c}[^{1}S_0^{[8]},^{3}P_J^{[8]}]+c\bar{c}[^{3}S_1^{[8]}]$ exhibits the largest SDCs, significantly exceeding those of the CS+CS process. Such large SDCs are expected to compensate for the relatively small CO LDMEs, thereby enabling contributions comparable to the pure CS process. Among the CS+CO channels, the dominant contribution arises from $e^{+}e^{-} \to c\bar{c}[^{3}S_1^{[1]}]+c\bar{c}[^{1}P_1^{[8]}]+g$, which involves only one CO state and is thus also expected to yield sizable effects.

\subsection{Cross Sections}
\begin{table*}[htb]
\begin{center}
\caption{Inclusive $J/\psi+\eta_c$ production cross sections (in fb) in $e^{+}e^{-}$ annihilation as a function of the center-of-mass energy. The calculation adopts $\mu_r=2m_c$ with $m_c=1.5$ GeV. ``CS'' and ``NR'' correspond to the CS and NRQCD predictions, respectively, and the error bands represent the LDME uncertainties.}
\label{tab: cross sections}
\begin{tabular}{|c|cc|cc|cc|cc|cccc}
\hline
\multirow{2}{*}{$\sqrt{s}$}  & \multicolumn{2}{c|}{set 1} & \multicolumn{2}{c|}{set 2} & \multicolumn{2}{c|}{set 3}  & \multicolumn{2}{c|}{set 4}\\
~& $\sigma_{\textrm{CS}}$ & $\sigma_{\textrm{NR}}$ & $\sigma_{\textrm{CS}}$ & $\sigma_{\textrm{NR}}$ & $\sigma_{\textrm{CS}}$ & $\sigma_{\textrm{NR}}$ & $\sigma_{\textrm{CS}}$ & $\sigma_{\textrm{NR}}$ \\ \hline
10.52 GeV & 16.12 & $15.81^{+0.46}_{-0.42}$ & 12.45 & $15.64^{+0.89}_{-0.83}$ & 12.45 & $15.93^{+2.15}_{-1.81}$ & 12.45 & $15.02^{+0.42}_{-0.41}$ \\ \hline
$\Upsilon(4S)$ & 15.58 & $15.27^{+0.45}_{-0.41}$ & 12.03 & $15.18^{+0.88}_{-0.82}$ & 12.03 & $15.47^{+2.12}_{-1.79}$ & 12.03 & $14.58^{+0.41}_{-0.41}$ \\ \hline
$\Upsilon(5S)$ & 13.06 & $12.78^{+0.43}_{-0.39}$ & 10.08 & $13.04^{+0.82}_{-0.76}$ & 10.08 & $13.31^{+1.97}_{-1.67}$ & 10.08 & $12.50^{+0.39}_{-0.38}$ \\ \hline
15 GeV & 1.626 & $1.608^{+0.17}_{-0.16}$ & 1.256 & $2.472^{+0.31}_{-0.29}$ & 1.256 & $2.595^{+0.75}_{-0.64}$ & 1.256 & $2.141^{+0.15}_{-0.14}$ \\ \hline
20 GeV & 0.232 & $0.258^{+0.06}_{-0.06}$ & 0.179 & $0.656^{+0.11}_{-0.11}$ & 0.179 & $0.710^{+0.28}_{-0.24}$ & 0.179 & $0.465^{+0.05}_{-0.05}$ \\ \hline
\end{tabular}
\end{center}
\end{table*}

\begin{figure}[!h]
\begin{center}
\hspace{0cm}\includegraphics[width=0.47\textwidth]{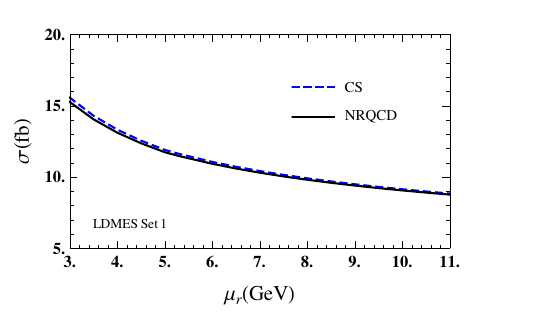}
\hspace{0cm}\includegraphics[width=0.47\textwidth]{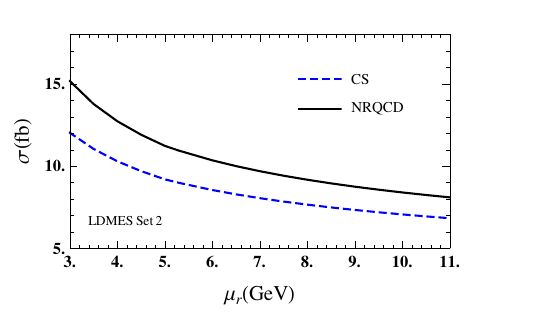}
\hspace{0cm}\includegraphics[width=0.47\textwidth]{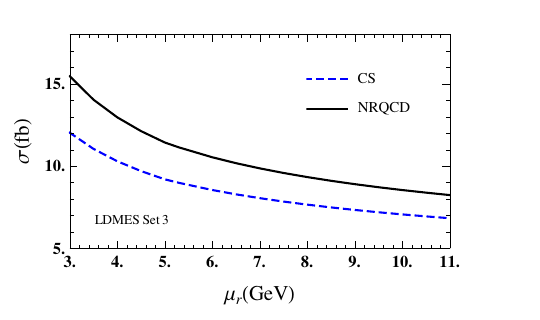}
\hspace{0cm}\includegraphics[width=0.47\textwidth]{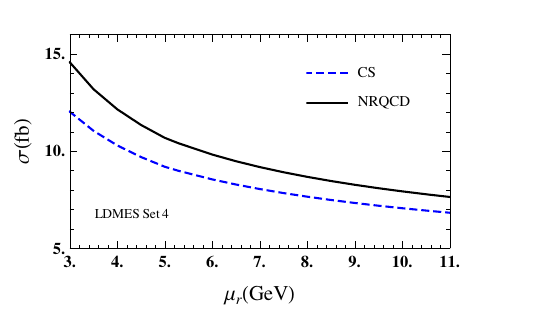}
\caption{\label{mu_running}
The cross sections for inclusive $J/\psi+\eta_c$ production in $e^{+}e^{-}$ annihilation as a function of the renormalization scale $\mu_r$. $\sqrt{s}=10.58$ GeV and $m_c=1.5$ GeV.}
\end{center}
\end{figure}

Table \ref{tab: cross sections} presents the NLO predictions for inclusive $J/\psi+\eta_c$ production in $e^+e^-$ annihilation within the NRQCD framework. The ``CS'' results include both the pure QCD contributions and the interference with QED diagrams, while the ``NR'' (NRQCD) predictions incorporate all relevant production channels listed in equation (\ref{channels}). As shown in the table, the CS and NRQCD predictions are in close agreement for LDME set 1. In contrast, for the other three LDME sets, the NRQCD predictions are significantly larger, with notable differences. For instance, at $\sqrt{s} = \Upsilon(4S)$, the NRQCD results enhance the CS predictions by about $20\%$--$30\%$; at $\sqrt{s} = 15$~GeV, they are about 2 times larger; and at $\sqrt{s} = 20$~GeV, they can reach up to $3$--$4$ times the CS values. To examine the dependence of the NLO predictions on the renormalization scale $\mu_r$, we take $\sqrt{s} = \Upsilon(4S)$ as a reference and present in figure \ref{mu_running} the $\mu_r$-dependence of the results for the different LDME sets. For set~1, the CS and NRQCD predictions become increasingly compatible as $\mu_r$ increases. In contrast, for the other three LDME sets, significant differences between the two frameworks persist over a broad range of $\mu_r$ values.

Note that in deriving the results in Table~\ref{tab: cross sections}, we have fixed $\alpha$ at $1/137$. However, for $\sqrt{s}=20$~GeV, if the running of $\alpha$ is taken into account and the scale is set to $20$~GeV, $\alpha$ increases from $1/137$ to about $1/129$, corresponding to an enhancement of about $6\%$. This leads to an increase of about $15\%$ for the CS+CS process and a uniform increase of about $13\%$ for the CS+CO and CO+CO processes.

We assess the experimental feasibility of observing the inclusive process $e^{+}e^{-} \to J/\psi+\eta_c+X$ using the reconstruction strategy adopted by the Belle Collaboration in their measurement of $e^{+}e^{-} \to \eta_c J/\psi$ \cite{Belle:2023gln}, performed as part of the search for the $T_{cc}$ state. In this approach, both the $J/\psi$ and the $\eta_c$ are reconstructed simultaneously. Specifically, the $J/\psi$ is identified via the dimuon decay channel $J/\psi \to \mu^+\mu^-$, while the $\eta_c$ is reconstructed through six hadronic decay channels: $p\bar{p}$, $p\bar{p}\pi^0$, $K_S^0 K^\pm \pi^\mp$, $K^+ K^- \pi^0$, $K^+ K^- K^+ K^-$, and $2(\pi^+\pi^-\pi^0)$. This reconstruction strategy can be applied to the inclusive $J/\psi+\eta_c$ process studied in this work. Following ref.~\cite{Belle:2023gln}, the relation between the number of signal events and the cross section is given by
\begin{eqnarray}
N_{\textrm{sig}}=\sigma\times\epsilon\mathcal{L}\mathcal{B}(J/\psi \to \mu^{+}\mu^{-})\mathcal{B}(\eta_c \to 6~\textrm{channels}),
\end{eqnarray}
where $N_{\text{sig}}$ denotes the number of signal events, $\epsilon$ is the reconstruction efficiency, and $\mathcal{L}$ is the integrated luminosity. The specific values of $\epsilon$, $\mathcal{L}$, and the product $\mathcal{B}(J/\psi \to \mu^{+}\mu^{-}) \, \mathcal{B}(\eta_c \to \text{6 channels})$ are listed in table 1 of ref.~\cite{Belle:2023gln}. Based on these inputs, at $\sqrt{s}=10.52$~GeV and at the $\Upsilon(5S)$ resonance, a few events are expected to be collected, whereas at the $\Upsilon(4S)$ resonance, approximately a dozen events can be accumulated. Given the extremely high integrated luminosity of \(50~\textrm{ab}^{-1}\) foreseen at SuperKEKB, the process under consideration is expected to accumulate several hundred to a thousand events near the \(\Upsilon(4S)\) resonance, thus indicating encouraging prospects for experimental observation.

\begin{figure}[!h]
\begin{center}
\hspace{0cm}\includegraphics[width=0.47\textwidth]{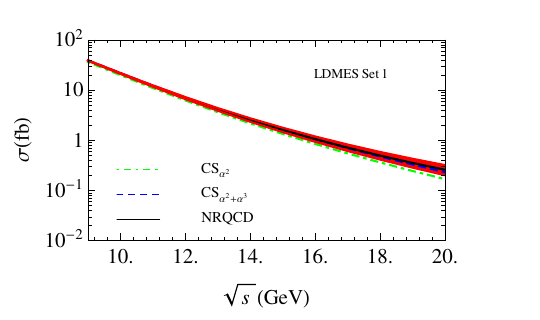}
\hspace{0cm}\includegraphics[width=0.47\textwidth]{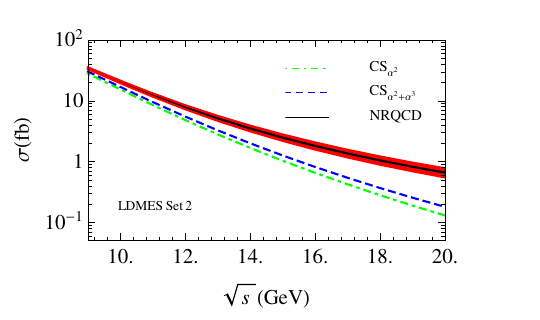}
\hspace{0cm}\includegraphics[width=0.47\textwidth]{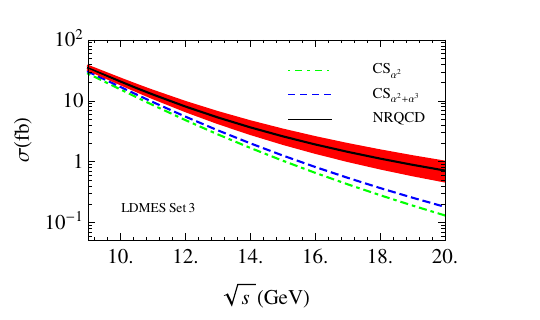}
\hspace{0cm}\includegraphics[width=0.47\textwidth]{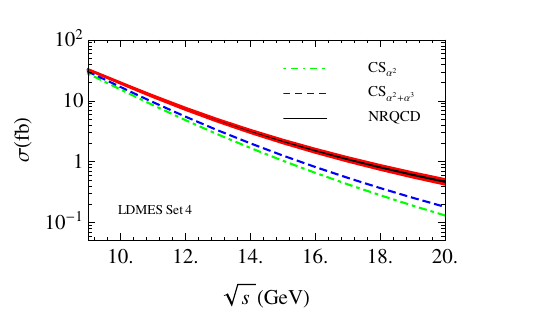}
\caption{\label{sqrts_running}
The cross sections for inclusive $J/\psi+\eta_c$ production in $e^{+}e^{-}$ annihilation as a function of $\sqrt{s}$. The renormalization scale is fixed at $\mu_r=2m_c$ with $m_c=1.5$ GeV. The error bands reflect the uncertainties in the LDMEs.}
\end{center}
\end{figure}

In the following, we investigate the dependence of the inclusive $e^{+}e^{-} \to J/\psi+\eta_c+X$ process on the collision energy. Figure \ref{sqrts_running} presents the cross-section predictions as a function of $\sqrt{s}$ obtained with different LDME sets, with error bands reflecting the LDME uncertainties. In the figure, $\sigma_{\alpha^2}$ denotes the sum $\sigma^{(0)}_{\alpha^2}+\sigma^{(1)}_{\alpha^2}$ from equation (\ref{Eq. QED+QCD}), while $\sigma_{\alpha^2+\alpha^3}$ represents $\sigma^{(0)}_{\alpha^2}+\sigma^{(1)}_{\alpha^2}+\sigma^{(0)}_{\alpha^3}+\sigma^{(1)}_{\alpha^3}$. For the CS+CS process, as $\sqrt{s}$ increases, the interference between QED and QCD diagrams becomes increasingly important relative to the pure QCD contribution. At $\sqrt{s}=20$~GeV, this interference term enhances the NLO QCD results by more than $40\%$, which—compared with the $\sim 60\%$ enhancement from NNLO corrections—underscores its significant theoretical value. For the NRQCD predictions, those obtained with set 1 LDMEs are very close to the CS results, exhibiting a similar $\sim s^{-4}$ scaling, and become almost indistinguishable once the LDME uncertainties are included. In contrast, the other three LDME sets yield NRQCD predictions with a noticeably slower decrease, approximately $\sim s^{-3}$, as $\sqrt{s}$ increases.

Finally, we estimate the contributions from feed-down of higher excited states to the inclusive $J/\psi+\eta_c$ production process. For the $J/\psi+h_c$ and $\chi_c+\eta_c$ channels, the Born-level contributions necessitate associated photon emission due to $C$-parity conservation, and are therefore negligible. The $\chi_c+h_c$ channel, which involves two-step feed-down and has an intrinsically small cross section \cite{Braaten:2002fi}, is also neglected. We thus consider only the process $e^+e^- \to \psi(2S)+\eta_c+X$ as the source of feed-down contributions. Using the central values of the $\psi(2S)$ LDMEs from ref.~\cite{Gong:2012ug},
\begin{eqnarray}
\langle \mathcal O ^{\psi(2S)}(^3S_1^{[1]}) \rangle & = & 0.76~\textrm{GeV}^3,\nonumber \\
\langle \mathcal O ^{\psi(2S)}(^1S_0^{[8]}) \rangle & = & -0.012 \times 10^{-2}~\textrm{GeV}^3, \nonumber \\
\langle \mathcal O ^{\psi(2S)}(^3S_1^{[8]}) \rangle & = & 0.34 \times 10^{-2}~\textrm{GeV}^3, \nonumber \\
\frac{\langle \mathcal O ^{\psi(2S)}(^3P_0^{[8]}) \rangle}{m_c^2} & = & 0.42 \times 10^{-2}~\textrm{GeV}^3,
\end{eqnarray}
we obtain at $\sqrt{s} = \Upsilon(4S)$:
\begin{eqnarray}
\sigma^{\text{CS}}_{e^+e^- \to \psi(2S)+\eta_c+X} &= 9.58~\text{fb}, \nonumber \\
\sigma^{\text{NRQCD}}_{e^+e^- \to \psi(2S)+\eta_c+X} &= 10.85~\text{fb}.
\end{eqnarray}
Using the branching fraction $\mathcal{B}(\psi(2S) \to J/\psi + X) = 0.595$, the resulting feed-down contribution to the inclusive $J/\psi+\eta_c$ process amounts to approximately $6.5~\text{fb}$, thereby increasing the central-LDME prediction by about $40\%$. Accordingly, the previously estimated event yield increases by the same proportion.

We note that the present study of inclusive $J/\psi+\eta_c$ production in $e^+e^-$ annihilation does not include initial-state radiation (ISR) contributions. As a reference, in the inclusive single-$J/\psi$ production process, the ISR effect accounts for about $20\%$~\cite{Shao:2014rwa}. This value serves only as an indication of the potential magnitude of ISR effects, as the exact size for the process under consideration may differ. Nevertheless, we preliminarily estimate that ISR corrections are unlikely to alter our main conclusion regarding the significant role of the CO mechanism. This estimate, however, awaits a more definitive assessment based on a complete NLO calculation of ISR corrections for the individual channels. On the experimental side, we also suggest that future measurements of this process should suppress ISR backgrounds to enable a more precise comparison between NRQCD predictions and experimental data.

Furthermore, the open-charm associated production $e^+e^- \to J/\psi + \eta_c + c + \bar{c}$ is excluded from our calculations. Although it provides a sizable contribution to inclusive single-$J/\psi$ production ($e^+e^- \to J/\psi + c + \bar{c}$), it enters at a formally higher order in $\alpha_s$ compared with the processes under consideration. For instance, $e^{+}e^{-} \to c\bar{c}[{}^1S_0^{[8]}]+c\bar{c}[{}^3S_1^{[8]}]+c\bar{c}$ constitutes an NLO correction to $c\bar{c}[{}^1S_0^{[8]}]+c\bar{c}[{}^3S_1^{[8]}]+g$, thereby corresponding to NNLO for $c\bar{c}[{}^1S_0^{[8]}]+c\bar{c}[{}^3S_1^{[8]}]$. Consequently, it lies beyond the scope of our NLO analysis and is therefore omitted.

\section{Summary}\label{sum}

We perform an NLO study of the inclusive $e^{+}e^{-} \to J/\psi+\eta_c+X$ process within the NRQCD factorization framework, aiming to test NRQCD validity at low energies. QCD corrections substantially enhance the LO predictions: for the pure CS channel, the NLO corrections increase the cross section by roughly a factor of two, with an additional $10\%$--$20\%$ from QED interference at $\sqrt{s}=\Upsilon(4S)$. This QED contribution is non-negligible and becomes comparable to the NNLO corrections as $\sqrt{s}$ increases, reaching a similar level at $\sqrt{s}=20$~GeV. For the CO channels, the QCD corrections enhance the LO results by about a factor of three. Using four LDME sets, one set yields NRQCD predictions close to CS results with a $\sim s^{-4}$ scaling, while the other three give larger predictions with a softer $\sim s^{-3}$ dependence. This marked difference provides a unique probe of double-charmonium production at low energies. With the central LDME values, the inclusion of $\psi(2S)$ feed-down increases the NRQCD prediction by about $40\%$ at $\Upsilon(4S)$. With Belle's reconstruction strategy for the $\eta_c J/\psi$ pair, the process is found to have promising potential for observation. However, we note that the present calculation does not include ISR corrections or open-charm associated production; these are left for future investigation.

\acknowledgments
We are grateful to Dr. Hong Fei Zhang for valuable discussions. This work was partially supported by the National Natural Science Foundation of China (Grant No. 12065006), the Guizhou Provincial Top-Quality Course (Grant No. 2024JKXX0048), and the Science and Technology Project of Guizhou Province (Grant No. CXTD[2025]030).\\

\providecommand{\href}[2]{#2}\begingroup\raggedright

\end{document}